\documentclass[conference]{IEEEtran}
\IEEEoverridecommandlockouts
\usepackage{cite}
\usepackage{amsmath,amssymb,amsfonts}
\usepackage{graphicx,subcaption}
\usepackage{algorithm, algorithmic}
\usepackage{textcomp}
\usepackage{xcolor}
\usepackage{epsfig}
\usepackage{epstopdf}
\usepackage{balance}

\def\BibTeX{{\rm B\kern-.05em{\sc i\kern-.025em b}\kern-.08em
    T\kern-.1667em\lower.7ex\hbox{E}\kern-.125emX}}
\begin{document}

\title{Measurement-based Characterization of Physical Layer Security for RIS-assisted Wireless Systems
}

\author{\IEEEauthorblockN{
Samed Keşir\textsuperscript{$\ast$,$\circ$},
Sefa Kayraklık\textsuperscript{$\ast$,$\bullet$},
İbrahim Hökelek\textsuperscript{$\ast$},
Ali Emre Pusane\textsuperscript{$\circ$},
Ertugrul Basar\textsuperscript{$\bullet$},
Ali Görçin\textsuperscript{$\ast$,$\diamond$}}
	\IEEEauthorblockA{
	\textsuperscript{$\ast$} Communications and Signal Processing Research (HISAR) Lab., TUBITAK BILGEM, Kocaeli, Turkey\\
    \textsuperscript{$\circ$} Department of Electrical and Electronics Engineering, Boğaziçi University, Istanbul, Turkey \\	
	\textsuperscript{$\bullet$} CoreLab, Department of Electrical and Electronics Engineering, Koç University, Istanbul, Turkey \\
    \textsuperscript{$\diamond$} Department of Electronics and Communication Engineering, Yildiz Technical University, Istanbul, Turkey\\
		Email:
		\{samed.kesir, sefa.kayraklik, ibrahim.hokelek\}@tubitak.gov.tr,\\
		ali.pusane@boun.edu.tr,
		ebasar@ku.edu.tr,
		agorcin@yildiz.edu.tr
		\vspace{-3ex}}
}
\maketitle

\begin{abstract}
There have been recently many studies demonstrating that the performance of wireless communication systems can be significantly improved by a reconfigurable intelligent surface (RIS), which is an attractive technology due to its low power requirement and low complexity. This paper presents a measurement-based characterization of RISs for providing physical layer security, where the transmitter (Alice), the intended user (Bob), and the eavesdropper (Eve) are deployed in an indoor environment. Each user is equipped with a software-defined radio connected to a horn antenna. The phase shifts of reflecting elements are software controlled to collaboratively determine the amount of received signal power at the locations of Bob and Eve in such a way that the secrecy capacity is aimed to be maximized. An iterative method is utilized to configure a Greenerwave RIS prototype consisting of 76 passive reflecting elements. Computer simulation and measurement results demonstrate that an RIS can be an effective tool to significantly increase the secrecy capacity between Bob and Eve.
\end{abstract}


\section{Introduction}
A reconfigurable intelligent surface (RIS) is a metasurface consisting of passive reflecting elements, which are software controlled to collaboratively convey the reflected signals towards any desired location. It is an attractive technology, since encoding and decoding operations are not needed, and only minimal extra energy is used for soft programming of passive reflecting elements to determine the phase of the reflected signal. An RIS can be an effective tool to cope with high path loss and blockage problems in millimeter wave and terahertz frequencies. In addition to significantly increasing the coverage and communication quality, an RIS can also be utilized to provide information secrecy through physical layer security (PLS) for next-generation wireless communication systems.

According to Wyner’s wire-tap channel model~\cite{adw1975}, the information secrecy increases as the difference between signal-to-interference and noise ratio (SINR) values at the legitimate receiver and the eavesdropper becomes larger. Traditionally, there are two main instruments to increase the SINR difference: transmitting artificial noise and multiple input multiple output (MIMO) precoding. In the first instrument, artificial noise, which is designed using the channel information of the legitimate user, is added to the transmitted signal. The legitimate user is still able to decode the received signal while the eavesdropper cannot decode it reliably since its channel quality is significantly degraded due to the artificial noise. In the second instrument, the MIMO precoding matrix can be designed in such a way that the SINR difference between the legitimate user and the eavesdropper is sufficiently high to provide PLS~\cite{cywu2013}. In this approach, the MIMO precoding matrix has a similar functionality of the encryption key in traditional cryptographic systems so that secure communication between the transmitter and the legitimate receiver is ensured by pre-determined precoding matrices known only to legitimate receivers.

%In MIMO systems, the baseband signal is multiplied by the precoding matrix before sending it to the antennas, and the selection of this matrix determines the channel capacity between the transmitter and the receiver~\cite{cywu2013}.  

Although the majority of RIS studies focus on theoretical aspects, measurement-based studies demonstrating the advantages of RIS-assisted wireless communication systems have been recently gaining momentum~\cite{wang2021,9880837,xpei2021}. A radar cross section-based received power model for an RIS-assisted wireless communication system is presented in~\cite{wang2021}, where the proposed model is validated through the received power measurements. The path loss of an RIS-assisted communication channel is modeled in~\cite{9880837}, where the line-of-sight (LOS) path between the transmitter and receiver is blocked.  The results show that the modified models can reasonably predict the measurements. In~\cite{xpei2021}, indoor and outdoor measurements are performed using an RIS prototype consisting of 1100 controllable elements, where the authors propose an RIS configuration algorithm by exploiting the geometrical array properties and the strength of the signal received at the receiver. The considered RIS provides a 26 dB power gain for indoors compared to  27 dB and 17 dB power gains over 50 m and 500 m distances for outdoors, respectively. 

%the reflection coefficients are calculated not only a function of the control voltage imposed on the RIS element but also a function of the angles. Their model is validated through the received power measurements using an experimental platform. 

 %The authors extend the floating-intercept (FI) and close-in (CI) path loss models to incorporate the RIS which is configured for either specular or intelligent reflection. 
 
 % to enable beam-forming the impinging signal towards the receiver. 

%A video is streamed from a transmitter to a receiver over RF signals through the RIS whose reflection coefficients are controlled by adjusting the impedance of the patches to enable beamforming the impinging signal towards the receiver.

None of the above-mentioned measurement-based studies utilizes the RIS concept for improving PLS, where a theoretical framework has been presented by several papers~\cite{wxu2022,jzhang2021,9134962,gzhou2020}. For example, an RIS-assisted MIMO system is proposed for PLS assuming that the channel state information (CSI) of the legitimate user is perfectly known at the transmitter while it is not available for the eavesdropper~\cite{wxu2022}.
%The direct links from the transmitter to the legitimate user and eavesdropper are obstructed while only reflected links from the RIS are available. 
The phase shift matrix at the RIS and the transmit covariance matrix at the transmitter are iteratively designed to minimize the transmit power so that the remaining power can be utilized to add artificial noise to suppress the eavesdroppers. The stochastic geometry theory is utilized in~\cite{jzhang2021} to characterize the PLS of the downlink RIS-assisted transmission for randomly located users and a multi-antenna eavesdropper. Specifically, the authors derive an exact probability distribution of the received SINR and obtain analytical security performance measures. The secrecy outage probability is analytically derived in~\cite{9134962} when the legitimate user and the eavesdropper can receive signals not only directly from the source but also from RIS. The presented analytical model successfully predicts the gain in secrecy outage probability when the number of elements in RIS increases. In~\cite{gzhou2020}, the secrecy rate is maximized by jointly designing active beam-forming at the base station and passive beamforming at the RIS. 

%in which the optimization considers the transceiver hardware impairments. 

%when the legitimate user and the eavesdropper can receive signals both from the base station and RIS, in which the optimization considers the transceiver hardware impairments. 

%They present an iterative optimization approach, where the successive convex approximation is utilized to obtain the optimized active beamforming vector while the semidefinite program is used to solve the passive beamforming optimization sub-problem.

In this paper, we present a measurement-based characterization of PLS for an RIS-assisted wireless communication system.
%where the RIS is utilized to convey the reflected beam towards the intended user while eliminating the reflected beam to reach the unintended user. 
A Greenerwave RIS hardware prototype\cite{greenerwave}, which consists of 76 passive reflecting elements, together with a transmitter (Alice), an intended user (Bob), and an eavesdropper (Eve), are deployed within an indoor office environment. Alice transmits a single-tone sinusoidal signal of 100 kHz at 5.2 GHz using a software-defined radio (SDR) with a horn antenna and the reflected signal from the RIS is received by horn antennas connected to the SDRs of Bob and Eve. The phase shifts of reflecting elements are software controlled to collaboratively determine the amount of received signal power at the locations of Bob and Eve. Each reflecting element has four distinct states to be configured since its phase shift can be independently controlled as $0^\circ$ or $180^\circ$  with two PIN diodes corresponding to the horizontal and vertical polarizations.
It is prohibitively expensive to utilize an exhaustive search with $4^{76}$ possible RIS configurations. An iterative method that explores four possible states for each reflecting element individually to decide its best state and repeats this procedure until covering all elements sequentially is utilized as in\cite{popov2021experimental}. The objective function is set to maximize the secrecy capacity between Bob and Eve. The presented experiment results demonstrate that an RIS can be an effective tool to provide PLS such that the difference of the received signal powers between Bob and Eve is significantly increased by RIS. Our computer simulation experiments show the similar behavior verifying that an RIS can be effectively used for PLS.

%\textcolor{red}{TODO: Rest of the Paper}
\section{System Model}\label{sec:SystemModel}
% In this section, the signal transmission model of the RIS-assisted wireless communication system, the analytical model for physical layer security, and the iterative search algorithm to configure the RIS are presented. 

\begin{figure}[t]
    \includegraphics[width=1\linewidth]{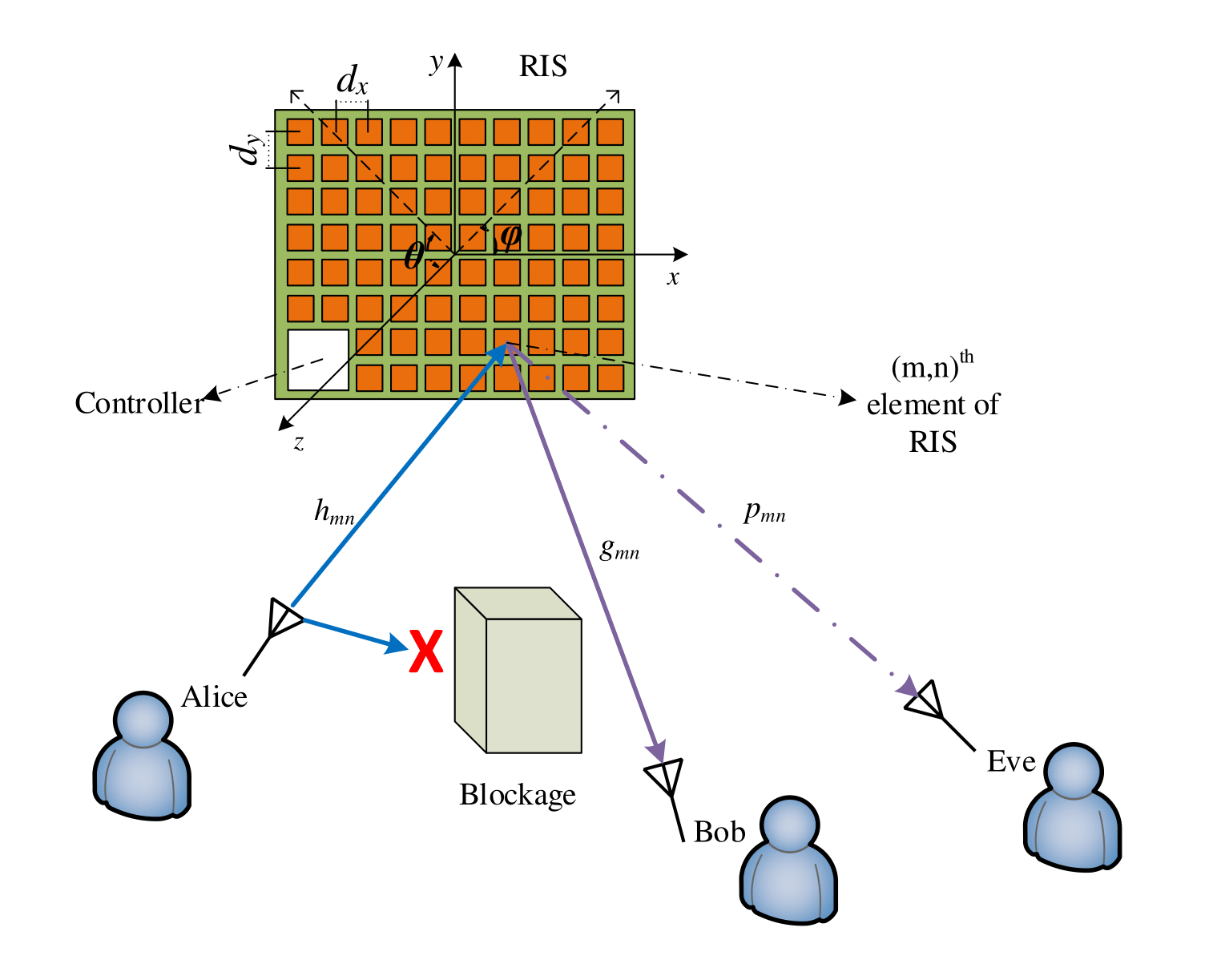}
    \caption{An RIS-assisted wireless communication system.} % TODO
    \label{fig:system_schema}
\end{figure}

The presented system model is shown in Fig. \ref{fig:system_schema}, where an RIS is located on the $xy$-plane and faces to the $+z$ direction. $\theta$ and $\phi$ represent the elevation and azimuth angles according to the $xy$-plane, respectively. The orange squares correspond to passive reflecting elements of the RIS that are controllable through the PIN diodes such that each element is responsible for adding either $0^\circ$ or $180^\circ$ phase delay. The controller of the RIS, which is depicted as a white square in the system model, alters the amount of voltage on the PIN diodes by turning the corresponding switches on or off. Alice (transmitter), Bob (intended user), and  Eve (eavesdropper) are located in such a way that there is no direct link from Alice to Bob and Eve (blockage) and they only communicate through the reflected signals from the RIS. 

% \subsection{Signal Transmission Model}

% The scattered electric field of the RIS will be described initially. Then, the received signals are described in terms of the transmitting and receiving channels.

When Alice transmits the complex baseband signal $x[k]$, it reaches Bob and Eve only through the RIS. The received signals at Bob ($y_b[k]$) and Eve ($y_e[k]$) can be expressed as 
\begin{equation}
\label{eq:rec_signal}
\begin{aligned}
y_b[k]=&  \sum_{m=0}^{M-1} \sum_{n=0}^{N-1} h_{mn} \Gamma_{mn} e^{j \phi_{mn}}  g_{mn} x[k] + n_b[k], \\
y_e[k]=&  \sum_{m=0}^{M-1} \sum_{n=0}^{N-1} h_{mn} \Gamma_{mn} e^{j \phi_{mn}}   p_{mn} x[k] + n_e[k],
\end{aligned}
\end{equation}
where $h_{mn}$ is the channel from Alice to the $\mathrm{(m,n)}^{\mathrm{th}}$ element of the RIS, while $g_{mn}$ and $p_{mn}$ represent the channels from the RIS to Bob and Eve, respectively. $M$ and $N$ denote the number of elements in $x$ and $y$ axis, respectively. $\Gamma_{mn} $ and $ \phi_{m n}$ are the reflection amplitude and phase of the $\mathrm{(m,n)}^{\mathrm{th}}$ element and $n_b[k]$ and $n_e[k]$ represent the noise components.

The scattering electric field of an RIS, which can be utilized to represent $h_{mn}$, $g_{mn}$, and $p_{mn}$ using the parameters of the system model, is defined in \cite{yang2016programmable} as a function of elevation and azimuth angles $(\vartheta, \varphi)$ and is given as
\begin{equation}
\label{eq:e-field}
\begin{aligned}
E(\vartheta, \varphi)= & \sum_{m=0}^{M-1} \sum_{n=0}^{N-1} A_{mn} e^{j \alpha_{mn}}   f_{mn}\left(\vartheta_{mn}, \varphi_{mn}\right)   \Gamma_{mn} e^{j \phi_{mn}} \\
& \times  f_{mn}(\vartheta, \varphi)   e^{j k_{0}\left(m d_{x} \sin \vartheta \cos \varphi+n d_{y} \sin \vartheta \sin \varphi\right)},
\end{aligned}
\end{equation}
where $A_{mn} $ and $ \alpha_{mn}$ are the illuminating amplitude and phase of the signal coming to the $\mathrm{(m,n)}^{\mathrm{th}}$ element. $\vartheta_{mn}$ and $\varphi_{mn}$ denote the elevation and azimuth angles of Alice relative to the $\mathrm{(m,n)}^{\mathrm{th}}$ element. The function $f_{mn}(.)$ represents the scattering pattern of the microstrip patch antenna for the $\mathrm{(m,n)}^{\mathrm{th}}$ element. It is assumed that Alice is in the near-field of the RIS, and hence, the scattering pattern is $f_{mn}(\vartheta_{mn}, \varphi_{mn})=\cos \vartheta_{mn}$ for the incident signal, while $f_{mn}(\vartheta, \varphi)=\cos \vartheta$  for the reflected signal assuming that Bob and Eve are in the far-field. $e^{j k_{0}\left(m d_{x} \sin \vartheta \cos \varphi+n d_{y} \sin \vartheta \sin \varphi\right)}$ is the array steering factor of the $\mathrm{(m,n)}^{\mathrm{th}}$ element, where $k_0$ is the free-space wave-number while $d_x$ and $d_y$ represent the distances between adjacent reflecting elements in $x$ and $y$ directions, respectively. 

Finally, $h_{mn}$, $g_{mn}$, and $p_{mn}$ in (\ref{eq:rec_signal}) can be expressed  using the components of (\ref{eq:e-field}) with the assumption of  free space path loss as 
\begin{equation}
\label{eq:channels}
\begin{aligned}
h_{mn} = & pl_{mn}(\lambda,R^a_{mn})\cos \vartheta_{mn},\\
g_{mn} = & pl_{mn}(\lambda,R^b_{mn}) \cos \vartheta_b\\
&\times e^{j k_0\left(m d_x \sin \vartheta_b \cos\varphi_b+n d_y \sin \vartheta_b \sin \varphi_b\right)}, \\
p_{mn} = & pl_{mn}(\lambda,R^e_{mn})\cos \vartheta_e\\
&\times e^{j k_0\left(m d_x \sin \vartheta_e \cos\varphi_e+n d_y \sin \vartheta_e \sin \varphi_e\right)},
\end{aligned}
\end{equation}
% Since the each element of the utilized RIS has 2 passive devices, $\Gamma_{mn} = 1$ and $\phi_{mn}$ takes one of the values $0^\circ$ and $180^\circ$
where $pl(\lambda,R)= \frac{\lambda}{4  \pi  R}  e^{-j 2 \pi R / \lambda}$ is the path loss function\cite{pathloss}, ($\vartheta_b$, $\varphi_b$) and ($\vartheta_e$, $\varphi_e$) are the elevation and azimuth angles of Bob and Eve, respectively. $R^{a}_{mn}$, $R^b_{mn}$, and $R^e_{mn}$ are the distances of Alice, Bob, and Eve to the $\mathrm{(m,n)}^{\mathrm{th}}$ element of RIS, respectively, and $\lambda$ is the wavelength of the input signal $x[k]$.

\section{RIS-assisted Physical Layer Security}\label{sec:RIS-PLS}
In this section, we present the secrecy capacity of our RIS-assisted wireless communication system in terms of the received signal powers of Bob and Eve according to the system model given in Section \ref{sec:SystemModel}. Secondly, an iterative algorithm aiming to maximize the secrecy capacity through configuring the phases of the RIS elements is presented.

The secrecy capacity between Bob and Eve can be expressed as 

\begin{equation}
\begin{aligned}
C_s &= \max \left(\log _2\left(1+\frac{S_b}{N_b}\right)-\log _2\left(1+\frac{S_e}{N_e}\right),0\right) \\
 &= \max \left(\log _2\left(\frac{S_b+N_b}{N_b} \middle/  \frac{S_e+N_e}{N_e}\right),0\right)\\
 &= \max \left(\log _2\left(\frac{S_b+N_b}{S_e+N_e}\right) + \log _2\left(\frac{N_e}{N_b}\right),0\right),
%  &= \max \Big(\log_2\left(S_b+N_b\right)-\log _2\left(S_e+N_e\right)\\ 
% &\hspace{2cm} +\log _2(N_e) - \log _2(N_b),0 \Big)
\end{aligned}
\end{equation}
where ($S_b$, $N_b$) and ($S_e$, $N_e$) are the signal and noise powers pairs at Bob and Eve, respectively. The secrecy capacity can be simplified in terms of the received signal and noise powers as
\begin{equation}
\label{eq:secrecy_capacity}
    C_s = \max \left(\log _2(P_b)-\log _2(P_e) + \log _2(N_e) - \log _2(N_b),0\right),
\end{equation}
where the received signal powers $P_b$ and $P_e$ at Bob and Eve.
% over $K$ many samples can be calculated as
% \begin{equation}
% \begin{aligned}
%     P_b &=  \frac{1}{K} \sum_{k=0}^{K} |y_b[k] | ^2 \\
%     P_e &=  \frac{1}{K} \sum_{k=0}^{K} |y_e[k] | ^2 .
% \end{aligned}
% \end{equation}
The received signals $y_b[k]$ and $y_e[k]$ given in (\ref{eq:rec_signal}) can be simply expressed using the channels in the vector forms as
\begin{equation}
\begin{aligned}
     y_b[k] &= \left(\mathbf{g}^H\mathbf{\Theta}\mathbf{h}\right)x[k] + n_b[k],\\
     y_e[k] &= \left(\mathbf{p}^H\mathbf{\Theta}\mathbf{h}\right)x[k] + n_e[k],
\end{aligned}
\end{equation}
where 
\begin{equation}
\begin{aligned}
    \mathbf{h} &= [h_{11}, h_{12}, ..., h_{mn}, ..., h_{MN}] \in \mathbb{C}^{MN \times 1},\\
    \mathbf{g} &= [g_{11}, g_{12}, ..., g_{mn}, ..., g_{MN}] \in \mathbb{C}^{MN \times 1},\\
    \mathbf{p} &= [p_{11}, p_{12}, ..., p_{mn}, ..., p_{MN}] \in \mathbb{C}^{MN \times 1},\\
    \mathbf{\Theta} &= \mathrm{diag}\{e^{j\phi_{11}}, e^{j\phi_{12}}, ..., e^{j\phi_{mn}}, ..., e^{j\phi_{MN}} \} \in \mathbb{C}^{MN \times MN}.
\end{aligned}
\end{equation}
Assuming that $n_b[k]$ and $n_e[k]$ are white Gaussian noise components, $P_b$ and $P_e$ in (\ref{eq:secrecy_capacity}) can be expressed as 
\begin{equation}
\label{eq:power_bobeve}
\begin{aligned}
    P_b &= \left|\mathbf{g}^H\mathbf{\Theta} \mathbf{h}\right|^2 P_{a} + N_b, \\
    P_e &= \left|\mathbf{p}^H\mathbf{\Theta} \mathbf{h}\right|^2 P_{a} + N_e,
\end{aligned}
\end{equation}
where $P_{a}$ is the transmitted power from Alice.

When (\ref{eq:power_bobeve}) is substituted into (\ref{eq:secrecy_capacity}), even if all the channel information is available, it is quite complicated to calculate $\mathbf{\Theta}$ maximizing $C_s$ theoretically. Therefore, in practical scenarios, the measurements of the received signal powers can be directly used as in (\ref{eq:secrecy_capacity}) to calculate $C_s$. Then, the maximum secrecy capacity can be obtained as
% If all the channel information is available,  $C_s$ can be theoretically calculated by substituting Eqn. (\ref{eq:power_bobeve}) into Eqn. (\ref{eq:secrecy_capacity}). However, when the channel information is not available in practical scenarios, the measurements of the received signal powers can be directly used in Eqn. (\ref{eq:secrecy_capacity}) to calculate $C_s$. Then, the maximum secrecy capacity can be obtained as follows:
% Note that Eqn. (\ref{eq:secrecy_capacity}) and then, the secrecy capacity is maximized as: 
\begin{equation}
\label{eq:max}
\begin{aligned}
\mathbf{\Theta}^*=\arg &\max_{\phi_{mn}, \forall  mn} C_s\\
&\text{ s.t. } \phi_{mn} \in\left\{0^{\circ}, 180^{\circ}\right\}, \quad \forall  mn .
\end{aligned}
\end{equation}

\begin{algorithm}[h]
\caption{Iterative algorithm for configuring RIS elements} \label{alg:ite}
\begin{algorithmic}[1]
\renewcommand{\algorithmicrequire}{\textbf{Input:}}
\renewcommand{\algorithmicensure}{\textbf{Output:}}
\REQUIRE $P_b$, $P_e$, $M$, $N$, and $S$
\ENSURE  \textit{States} (a vector including the states of RIS elements)
\STATE initialize \textit{States} with $0^\circ$ phase shift
\STATE $C_{s,max}$ $\gets - \infty$
\STATE \textit{StepList} $\gets$ shuffle   $[1,M\times N]$
\FOR {$i \in $ \textit{StepList}}
\FOR {$j \in [1,S]$ }
\STATE configure $i$th element with state $j$
\STATE measure $P_b$ and $P_e$
\STATE calculate $C_s$ assuming that $N_b=N_e$
\IF {($C_s>C_{s,max}$)}
\STATE update $i$th element of \textit{States} with state $j$
\STATE  $C_{s,max}$ $\gets$ $C_s$
\ENDIF
\ENDFOR
\ENDFOR
\RETURN \textit{States}
\end{algorithmic}
\end{algorithm}

Since an exhaustive search is not practical, Algorithm \ref{alg:ite}, inspired from \cite{popov2021experimental},  is utilized to determine the RIS configuration. $States$ is a vector including the phase shifts of the RIS elements which are initially configured as $0^\circ$. Each element has one state, which can be any of $S$ distinct values. Algorithm \ref{alg:ite} applies $S$ distinct phase shifts to each element, which is chosen randomly and only once. The corresponding element of $States$ is updated using the phase shift of the selected element that maximizes $C_s$ between Bob and Eve. The same process is sequentially performed for each of the remaining elements until all the elements are considered. Finally, Algorithm \ref{alg:ite} returns $States$ as the RIS configuration aiming to maximize the secrecy capacity.

\section{Computer Simulations}
%In this section, initially, the simulation setup is explained, where the placements of the RIS, Alice, Bob, and Eve are mentioned and the RIS structure is explained. After that, the simulation results are presented.

%\subsection{Simulation Setup}
The simulation environment is generated in MATLAB R2022a according to the system model in Fig. \ref{fig:system_schema}, where the RIS is placed at the origin. Alice, Bob, and Eve are located in the simulation region of $10m \times 10m \times 10m$ such that Alice is in the near field of the RIS while both Bob and Eve are in the far field~\cite{farfield}. The RIS prototype used for the measurement experiments consists of 76 elements. Therefore, the simulation scenario includes the RIS having $76$ elements such that $M=8$ and $N=10$ based on the equations in Sections \ref{sec:SystemModel} and \ref{sec:RIS-PLS}. Note that the controller part of the RIS has 4 antenna-free cells as shown in Fig. \ref{fig:system_schema}. Furthermore, all the units are located on the front side of the RIS because the RIS is reflecting but not permeable. Algorithm \ref{alg:ite} is implemented for the performance evaluation of PLS. 
% g_{mn} = & \frac{\lambda}{4  \pi  R^b_{mn}}  e^{j 2 \pi R^b_{mn} / \lambda} \cos \vartheta_b\\
% &\times e^{j k_0\left(m d_x \sin \vartheta_b \cos\varphi_b+n d_y \sin \vartheta_b \sin \varphi_b\right)} \\
% p_{mn} = & \frac{\lambda}{4  \pi  R^e_{mn}}  e^{j 2 \pi R^e_{mn} / \lambda} \cos \vartheta_e\\
% &\times e^{j k_0\left(m d_x \sin \vartheta_e \cos\varphi_e+n d_y \sin \vartheta_e \sin \varphi_e\right)}
% \end{aligned}
% \end{equation} 

\begin{figure}[t]
\centering
\begin{minipage}[h]{0.49\linewidth}
    \centering
    \includegraphics[width=1\linewidth]
    {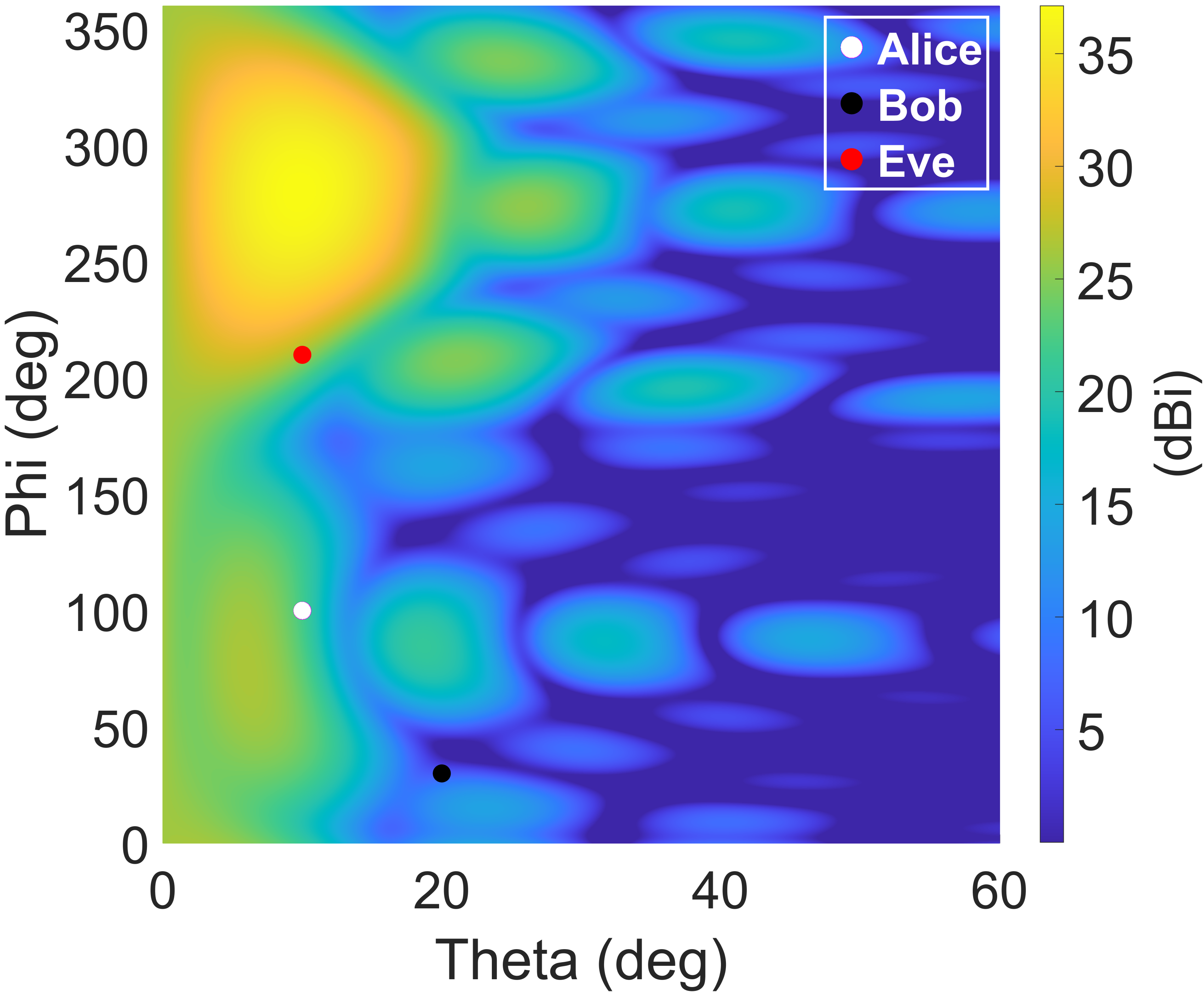}
    \subcaption{}
\end{minipage}
\begin{minipage}[h]{0.49\linewidth}
    \centering
    \includegraphics[width=1\linewidth]{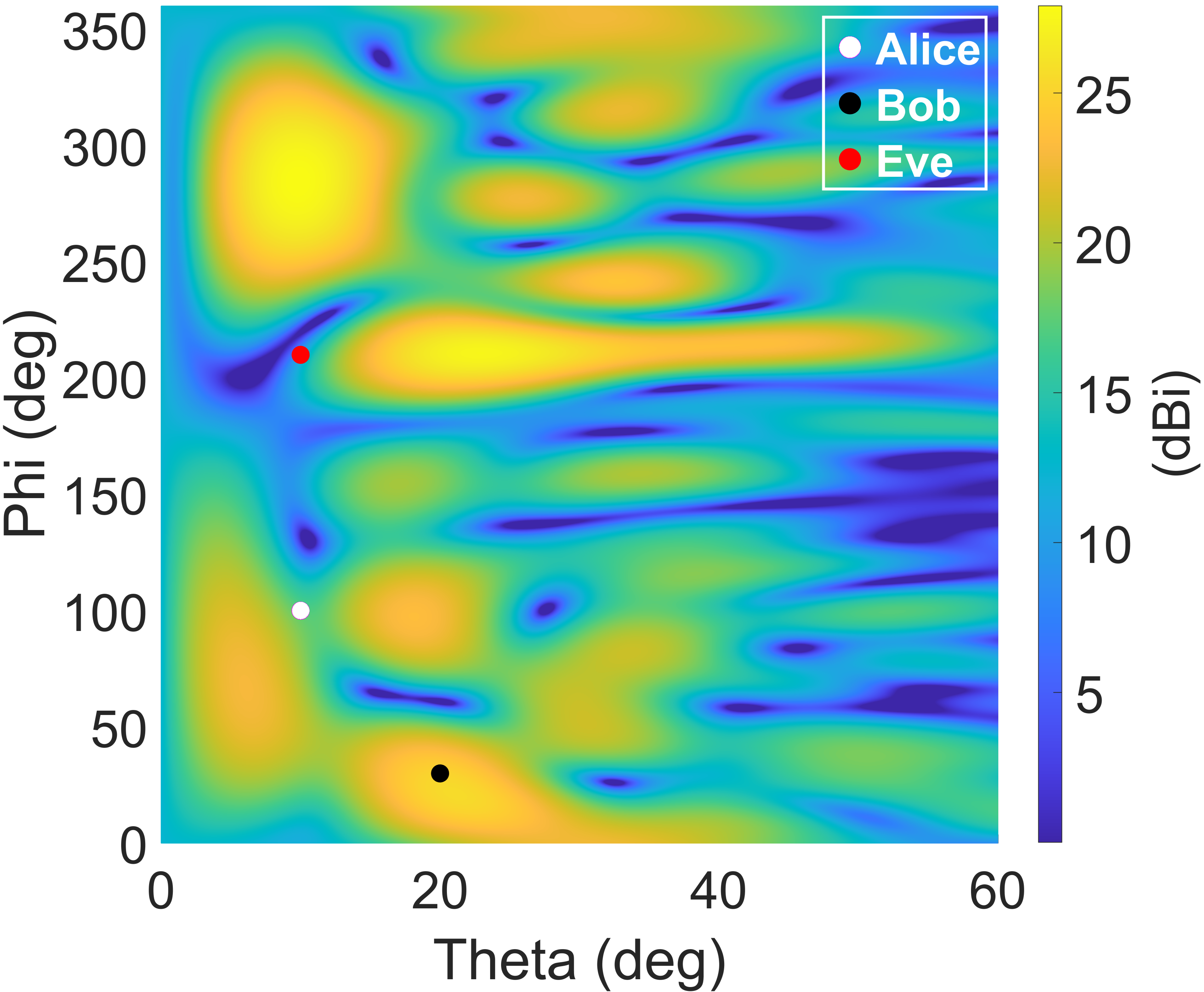}
    \subcaption{}
\end{minipage}

\caption[]{Scattered radiation pattern of the RIS (a) before the optimization and (b) after the optimization.}
\label{fig:rad_pattern} 
\end{figure}

%\subsection{Simulation Results}
Fig. \ref{fig:rad_pattern} (a) shows the scattered far-field radiation pattern of the RIS when there is no optimization while Fig. \ref{fig:rad_pattern} (b) shows it after the optimization, where the white, black, and red dots demonstrate the angular positions of Alice, Bob, and Eve, respectively. Note that the radiation level increases from $6.3$ dBi to $25.1$ dBi in the direction of Bob while it decreases from $25.9$ dBi to $10.1$ dBi in the direction of Eve and to illustrate the radiation pattern more clearly, all radiation levels below $0$ are visualized by the darkest blue color.
\begin{figure}[t]
    \centering
    \includegraphics[width=\linewidth]{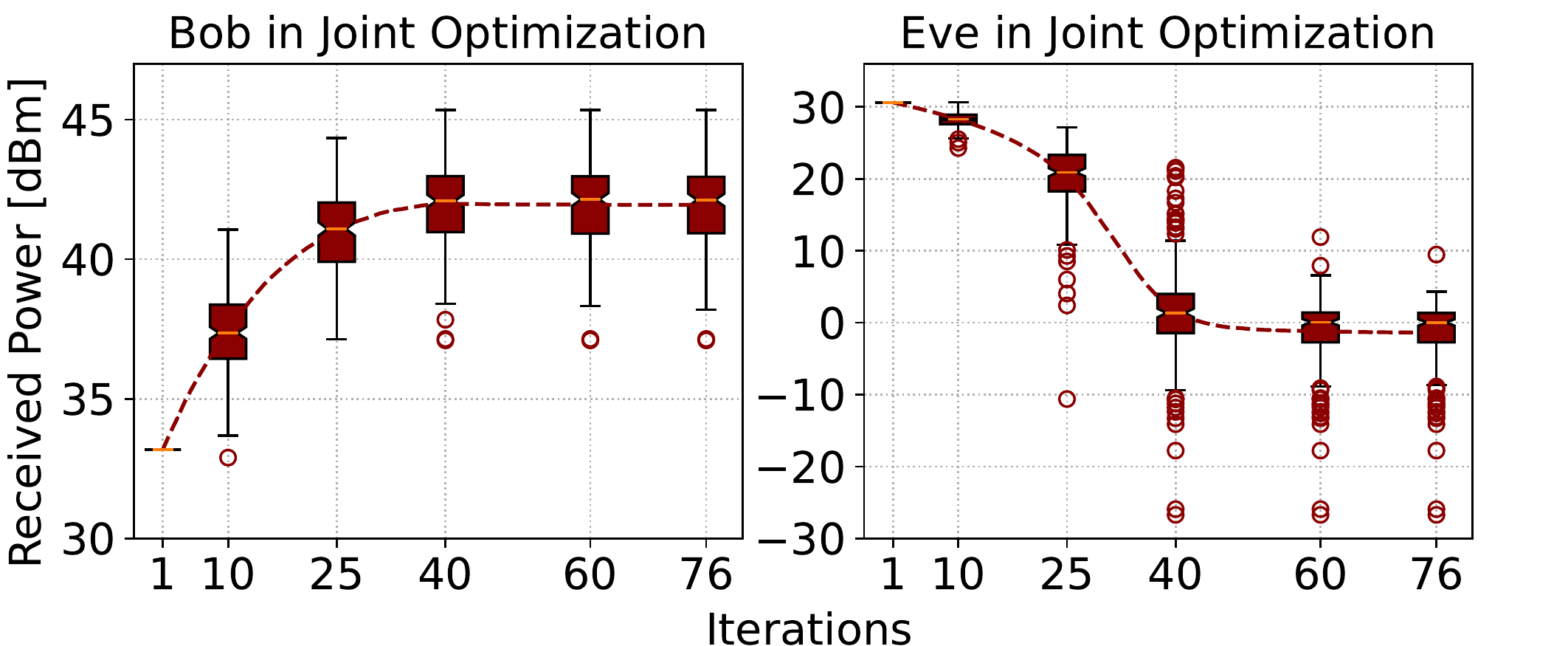}
    \caption{Simulation results of the received powers when the objective is to maximize the secrecy capacity.}
\label{fig:simStatsJoint}
\end{figure}

\begin{figure}[t]
\centering
\begin{minipage}[h]{0.47\linewidth}
    \centering
    \includegraphics[width=1\linewidth]{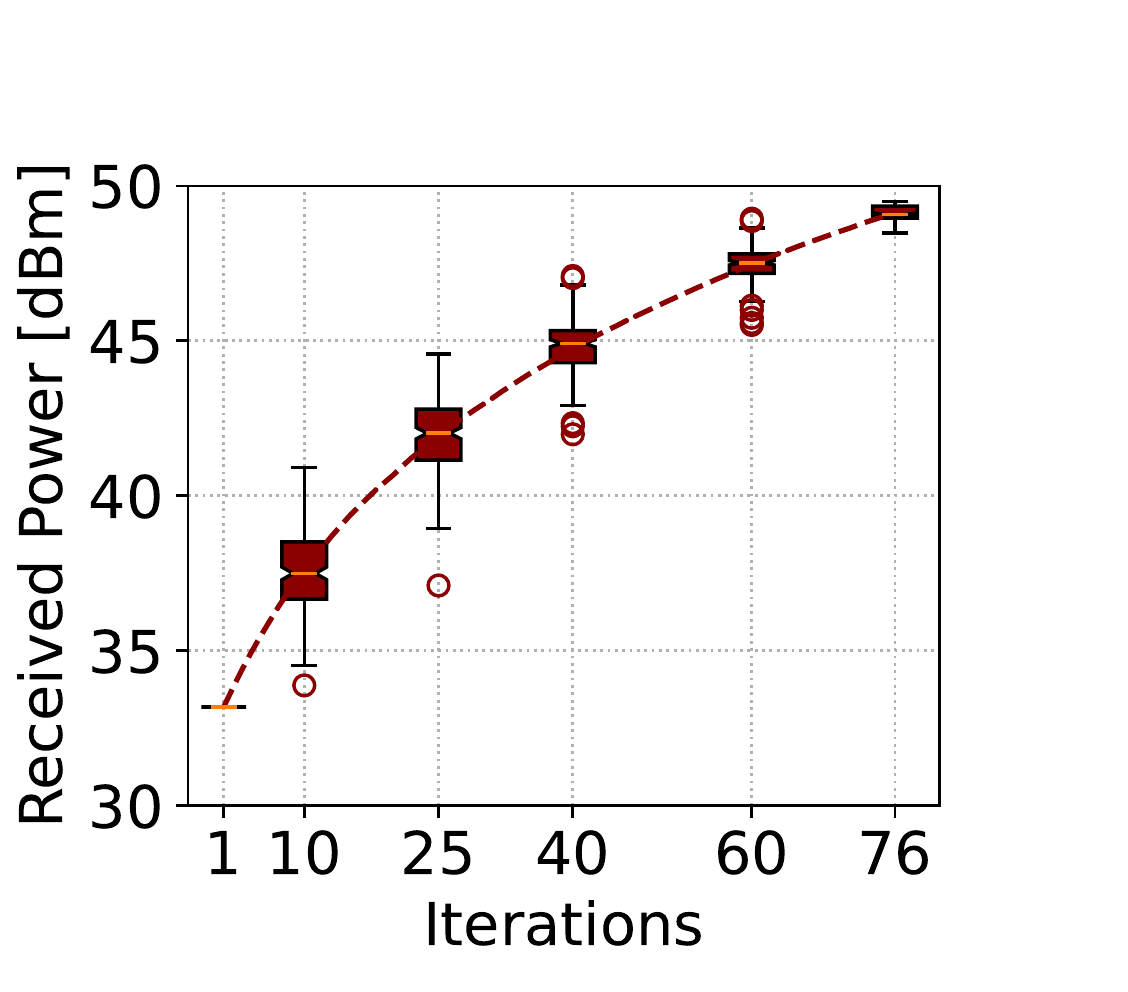}
    \subcaption{}
\end{minipage}
\begin{minipage}[h]{0.50\linewidth}
    \centering
    \includegraphics[width=1\linewidth]{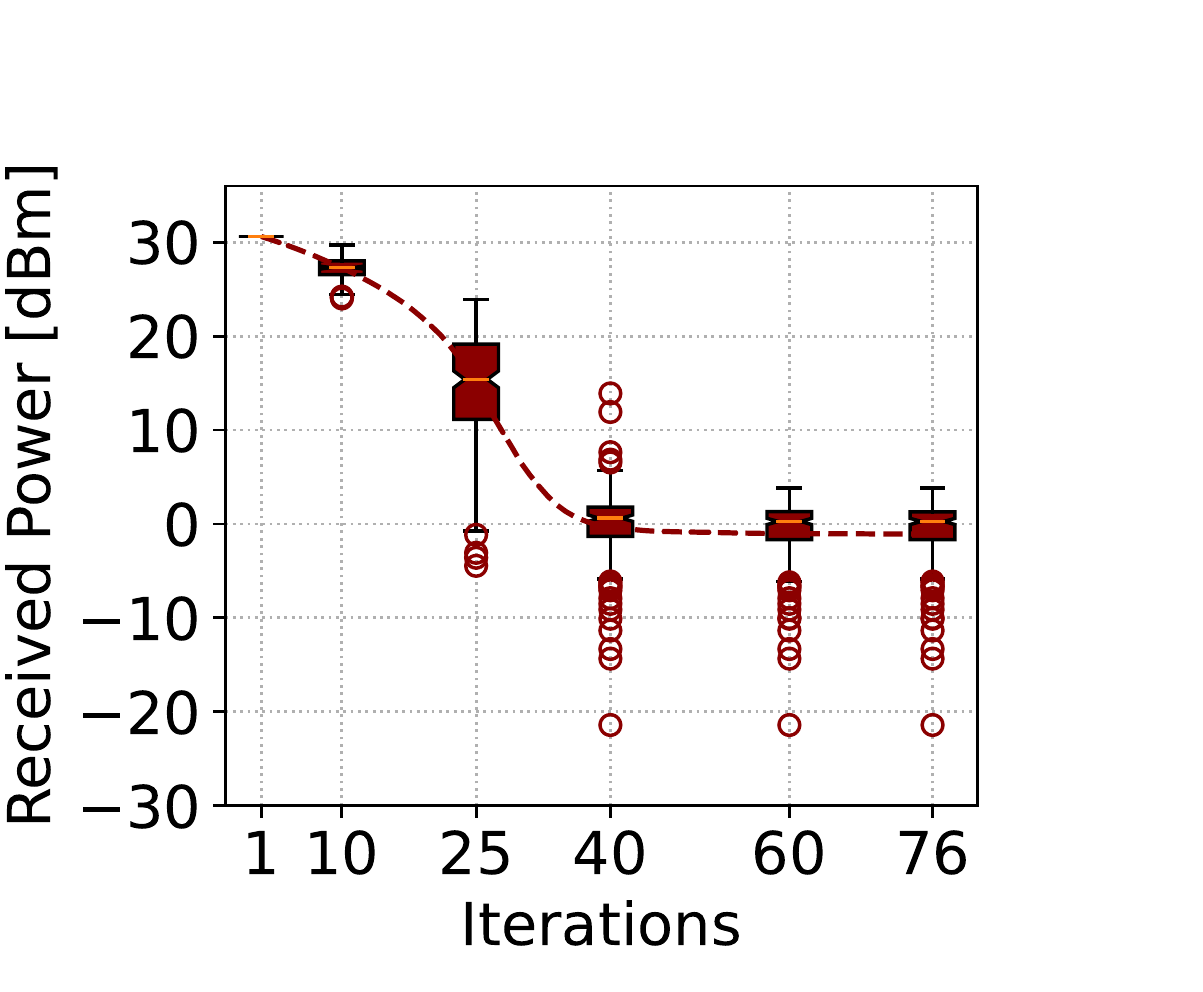}
    \subcaption{}
\end{minipage}
\caption[]{Simulation results of received power of (a) Bob when the objective is to maximize the received power, (b) Eve when the objective is to minimize the received power.}
\label{fig:simStats} 
\end{figure}

A set of computer simulations is performed to demonstrate the impact of the RIS configuration on the received powers of Bob and Eve, where Algorithm \ref{alg:ite} is used for configuring the RIS elements. Approximately $1000$ simulations of a practical scenario where Alice is in the near-field of the RIS and both Bob and Eve are in the far-field of the RIS, are conducted in a way that the order of elements in $StepList$ is randomly chosen for each simulation. The statistics of the received signal powers for Bob and Eve are illustrated in Fig. \ref{fig:simStatsJoint}, where the mean value at each iteration is shown as a dashed line. Furthermore, the boxplot shows the minimum, maximum, median, and the first and third quartiles of the received signal powers. The results demonstrate that the received power of Bob increases from $33$ dBm to $42$ dBm while the received power of Eve decreases from $30$ dBm to $0$ dBm on the average through the iterations and therefore, the secrecy capacity significantly increases. Fig. \ref{fig:simStats} (a) shows that the received signal power of Bob increases from $33$ dBm to $48$ dBm when the objective is set to maximize the received power of Bob only without considering the received power of Eve. Fig. \ref{fig:simStats} (b) illustrates that the received signal power of Eve decreases from $30$ dBm to $0$ dBm when the objective is set to minimize the received power of Eve only. Note that the received power of Eve fluctuates in both cases and therefore, this phenomenon causes fluctuations in Bob's received power as well when the objective is to maximize the secrecy capacity.  
% Eve fluctuation nedenleri Fig2'den açıklanabilir

\begin{figure}[t]
    \centering
    \includegraphics[width=\linewidth]{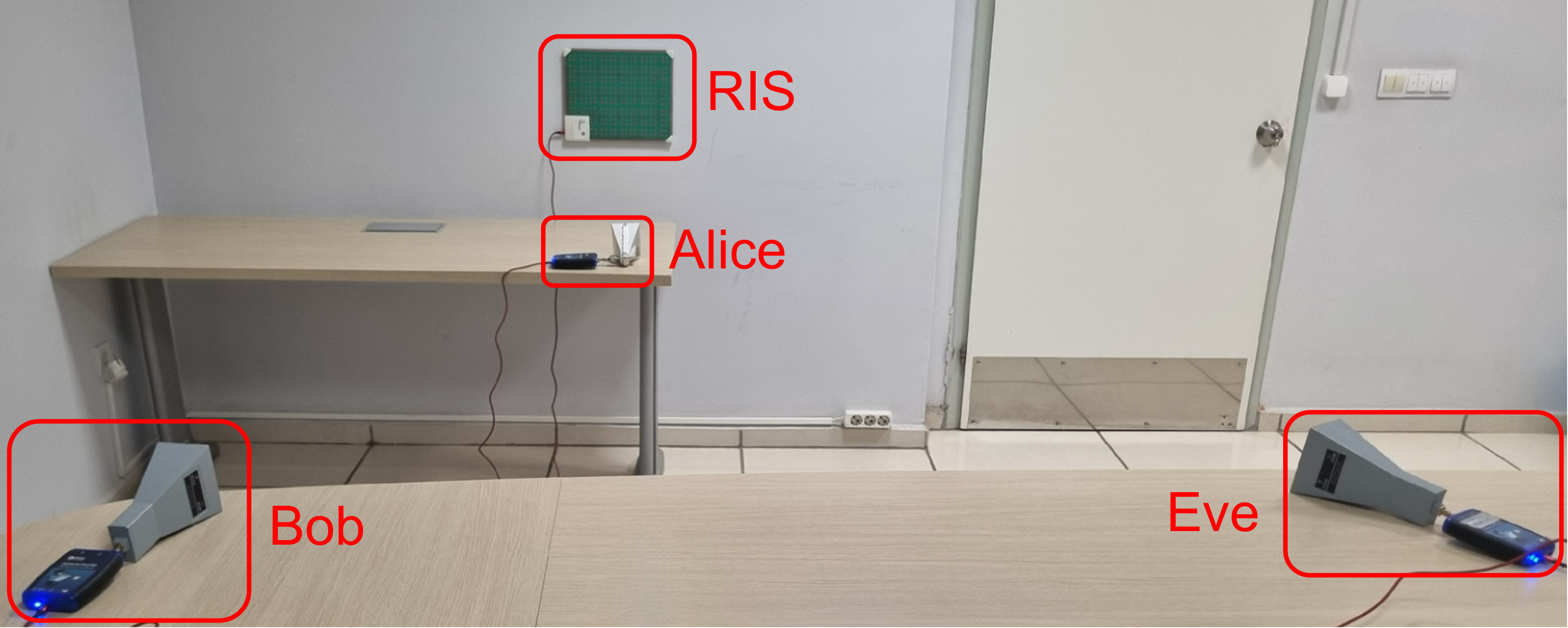}
    \caption{The measurement setup of the RIS-assisted wireless communication system.}
    \label{fig:schematic}
\end{figure}
\section{Practical Secrecy Measurements}\label{sec:MeasEnv}

The measurement setup of the RIS-aided wireless communication system, which is assembled in an indoor office environment by utilizing ADALM-PLUTO SDR modules to transmit and receive over-the-air signals, is illustrated in Fig. \ref{fig:schematic}. Alice is placed in front of the RIS, which is attached to the wall, while Bob and Eve are located further away from the RIS as shown in Fig. \ref{fig:schematic}. The Greenerwave RIS prototype\cite{greenerwave} is utilized to direct the beam from Alice to Bob while eliminating the beam against Eve by software-controlled phase shifts. The RIS comprises a uniform planar array form with $8\times 10$ reflecting elements, but the left bottom corner of the $2\times 2$ area is allocated for the controller, resulting in the total number of $76$ elements. The phase shift of each reflecting element can be independently controlled as $0^\circ$ or $180^\circ$ with two PIN diodes corresponding to the horizontal and vertical polarizations, ending up with four distinct states.

\begin{figure}[t]
    \centering 
    \includegraphics[width=0.7\linewidth]{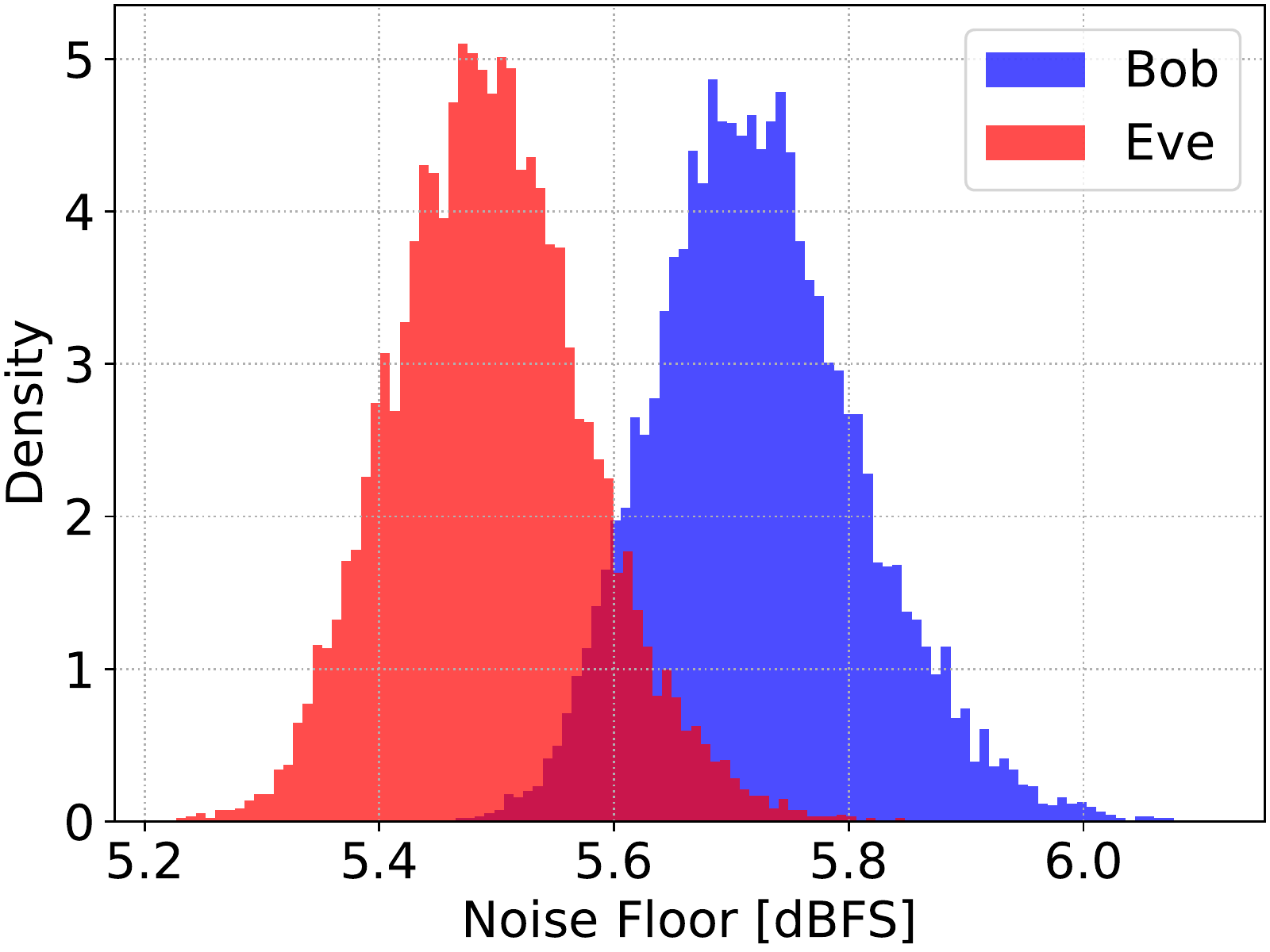}
    \caption{Noise Floor Densities of Bob and Eve.}\vspace{-0.1cm}
    \label{fig:noiseFloor}
\end{figure}

During the measurement campaign, a Python script controls the RIS by a Boolean array of $152$ elements corresponding to PIN diodes where $0$ and $1$ represent  $0^\circ$ and $180^\circ$ phase shifts, respectively. The sampling rate of the terminal SDRs is set to $1$ MHz while the buffer size of SDRs for both Bob and Eve, the number of time samples to compute the average received signal power, is selected as $10000$. Alice transmits a single-tone sinusoidal signal of $100$ kHz at $5.2$ GHz using a horn antenna facing the RIS, and the reflected signal from the RIS reaches the horn antennas of Bob and Eve. Each SDR captures the incoming signal at $5.2$ GHz and converts it to the complex baseband. The obtained complex baseband samples are depicted by integers in the range of $(-2047, 2048]$ since the SDR has a 12-bit analog to digital converter. Thus, the average power of the sampled received signal $y[k]$ can be calculated in the decibels relative to full scale (dBFS) as
\begin{equation}
    P_{dBFS}=10\log_{10} \left(\frac{1}{K}\sum_{k=1}^K |y[k]|^2\right),
\end{equation}
where $K$ stands for the total number of samples. Fig. \ref{fig:noiseFloor} shows the noise floor densities of Bob and Eve that are obtained from $11400$ measurements. The difference between the mean values of noise densities is approximately $0.228$ dB, which can be neglected. Therefore, the secrecy capacity defined in (\ref{eq:secrecy_capacity}) can be calculated by neglecting the difference in noise powers of Bob and Eve. % 0.22792804858154359 dB fark // 11400 ölçümün dağılımı
\begin{table}[!t]
\caption{The locations (x, y, z) of Alice, Bob, and Eve}
\label{tab:locs}
\begin{tabular}{l|lll}
     & Alice              & Bob                & Eve              \\ \hline 
Loc1 & (0.00, -0.35, 0.80)     & (-0.54, -0.35, 2.70) & (1.20,-0.35,2.60)  \\
Loc2 & (0.00, -0.35, 0.80)     & (1.20, -0.35, 2.60)  & (-0.54, -0.35, 2.70) \\
Loc3 & (-0.40, -0.35, 0.46) & (-0.10, -0.35, 2.60) & (0.80, -0.35, 3.80) \vspace{-12pt}
\end{tabular}
\end{table}

\begin{figure}[!t]
    \centering \vspace{-0.5cm}
    \includegraphics[width=\linewidth]{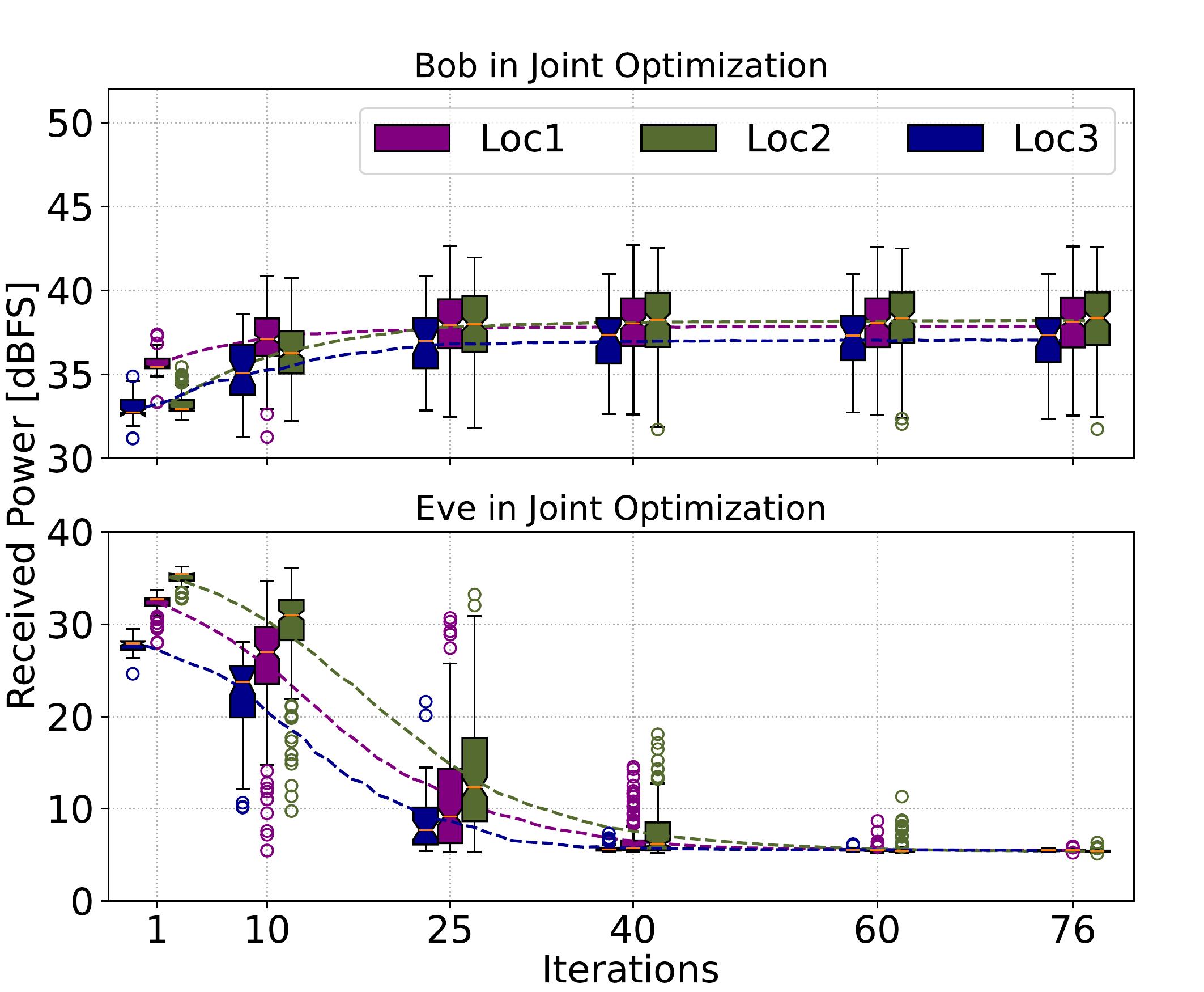}
    \caption{Measurement results of the received powers when the objective is to maximize the secrecy capacity.}\vspace{-0.65cm}
    \label{fig:measurementStatJoint}
\end{figure}
\begin{figure}[t]
    \centering
    \includegraphics[width=\linewidth]{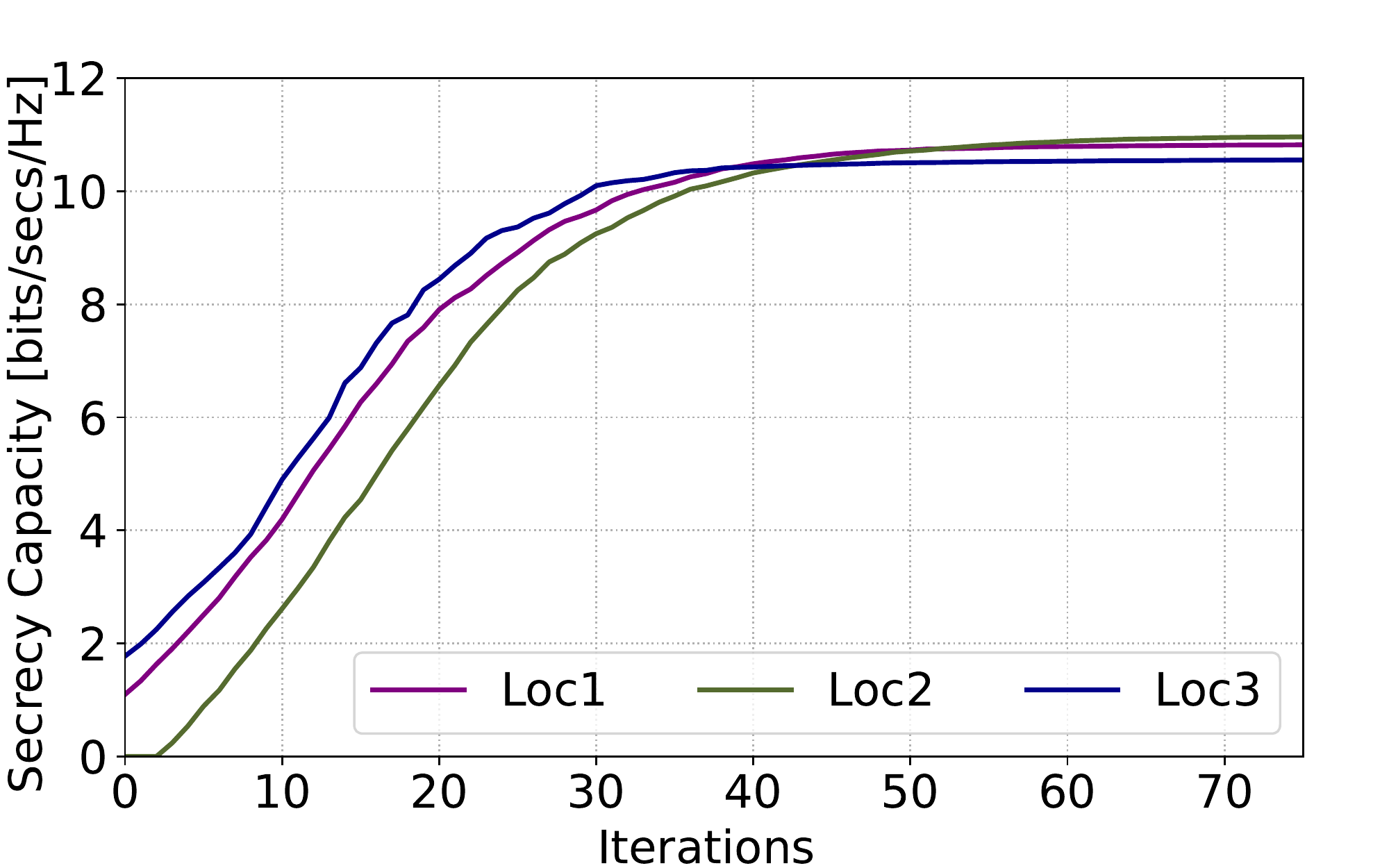}
    \caption{The average secrecy capacities for the measurement experiments.}\vspace{-0.70cm}
    \label{fig:measurementSec}
    
\end{figure}

The impact of the RIS on the received signal powers using Algorithm \ref{alg:ite} is demonstrated by performing a set of measurement experiments at three different locations in Table \ref{tab:locs}, where the RIS is placed at the origin. For each location, approximately $100$ measurements are conducted aiming to maximize the secrecy capacity, where the order of elements in $StepList$ is randomly chosen for each measurement. The received signal powers for Bob and Eve are illustrated in Fig. \ref{fig:measurementStatJoint} for three locations, where the mean values for each iteration are shown as dashed lines. The minimum, maximum, median, and the first and third quartiles of the received signal powers are depicted in the boxplots. The results indicate that the received power of Bob increases by approximately $5$ dB on average while the received power of Eve falls down to the noise floor. Note that the behaviors of both computer simulation and measurement results in Figs. \ref{fig:simStatsJoint} and \ref{fig:measurementStatJoint}, respectively, are similar, except Eve's received signal powers saturate at the noise floor in the measurement since the sensitivity of the PLUTO SDR is not sufficient to receive the signals below $5$ dBFS. The corresponding secrecy capacity results are demonstrated in Fig. \ref{fig:measurementSec}, where the secrecy capacities are significantly increased up to $11$ bps/Hz. For all locations, secrecy capacities increase rapidly until settling down around iteration $40$, which is consistent with Fig. \ref{fig:measurementStatJoint}.

\begin{figure}[t]
% \centering \vspace{-16pt}
\begin{minipage}[h]{\linewidth}
    \centering
    \includegraphics[width=1\linewidth]{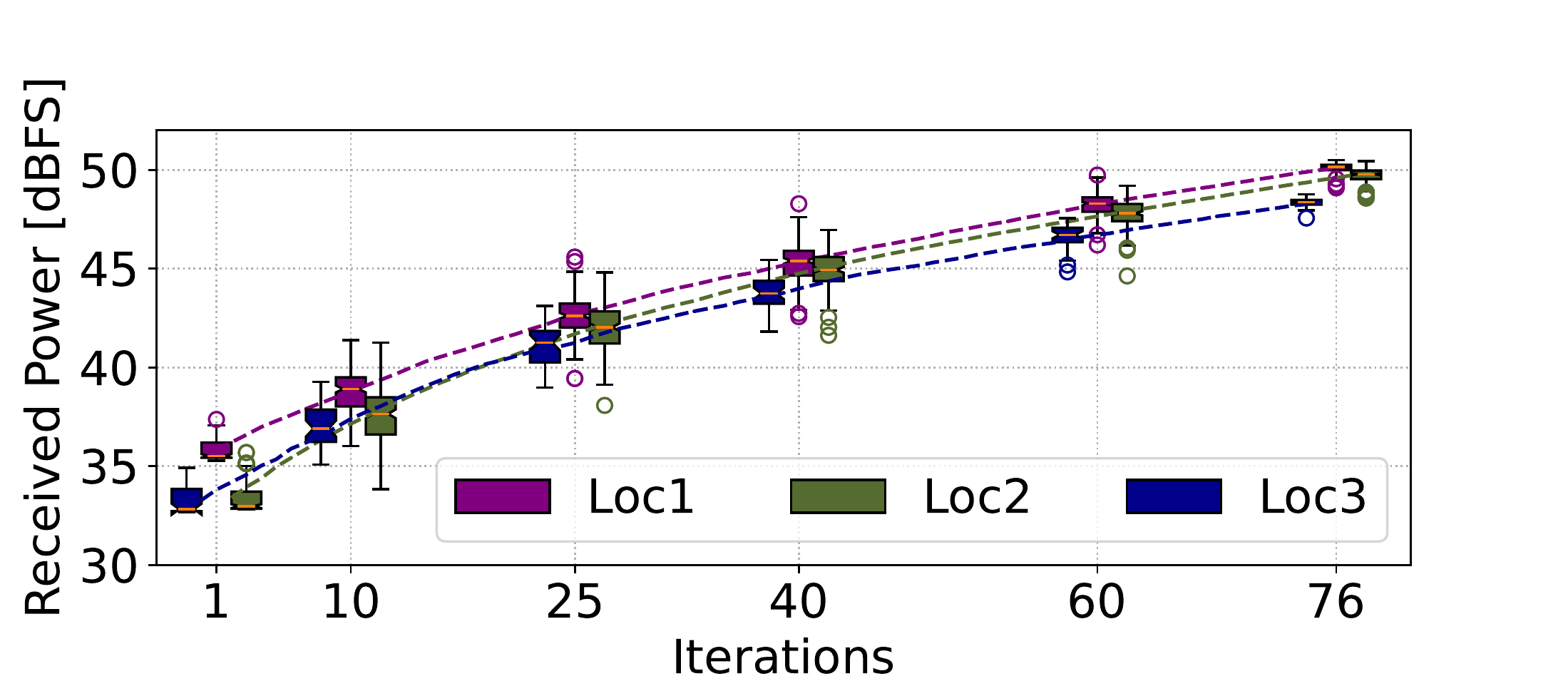}
    \subcaption{}
\end{minipage}
\begin{minipage}[h]{\linewidth}
    \centering \vspace{-2pt}
    \includegraphics[width=1\linewidth]{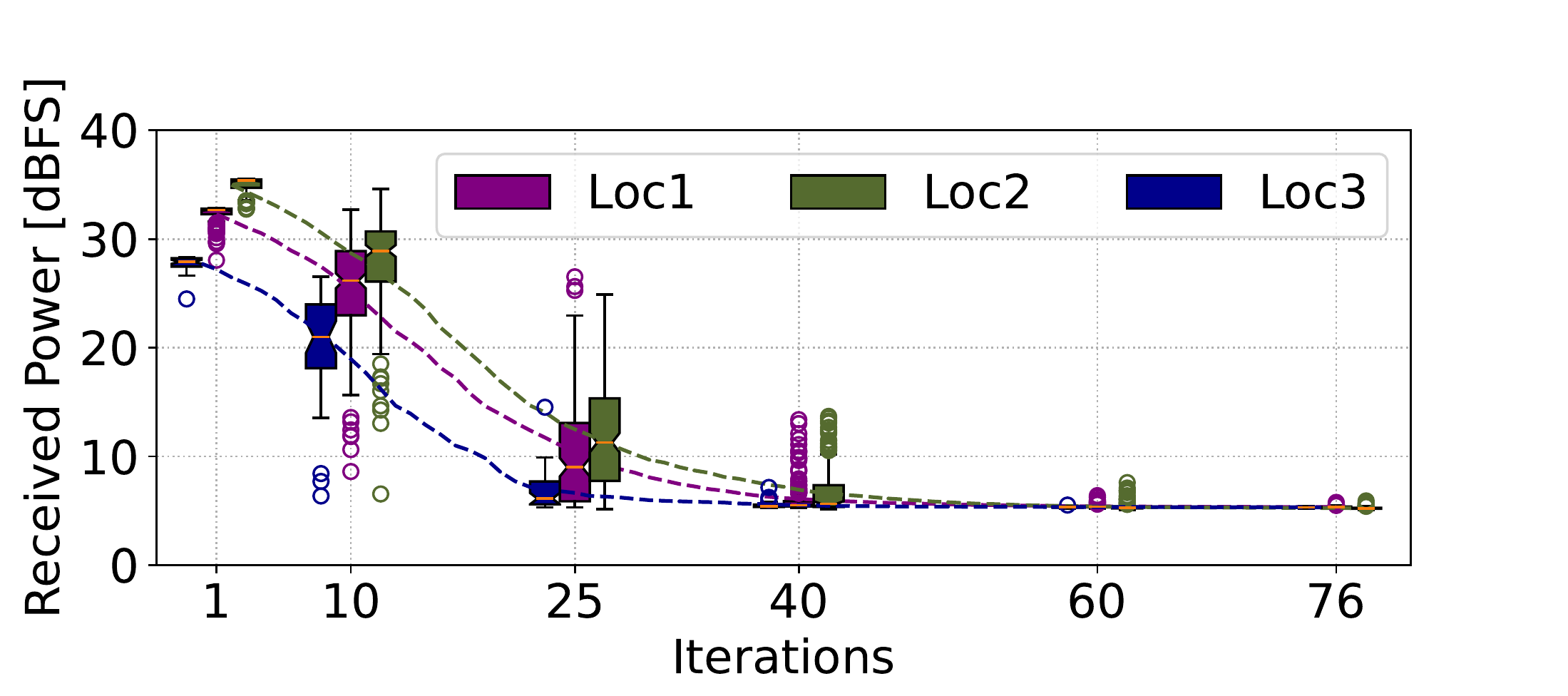}
    \subcaption{}
\end{minipage} \vspace{-1pt}
\caption[]{Measurement results of received power of (a) Bob when the objective is to maximize the received power, (b) Eve when the objective is to minimize the received power.}
\vspace{-0.5cm}
\label{fig:measurementStat} 
\end{figure}

The received signal power of Bob increases by $14$ dB on average when the objective is set to maximize the received power of Bob only as demonstrated in Fig. \ref{fig:measurementStat} (a) while the received signal power of Eve falls down to the noise floor when the objective is set to minimize the received power of Eve only as shown in Fig. \ref{fig:measurementStat} (b). We observe a similar behavior of the computer simulations such that the received power of Bob fluctuates when the objective is to maximize the secrecy capacity while it converges when the objective is to maximize Bob's received power only. As the scattered electric field in Fig. \ref{fig:rad_pattern} shows, there exist many regions with low signal powers and therefore, the iterative algorithm prioritizes to further decreasing Eve's received power to improve the secrecy capacity. 

\section{Conclusion}
\vspace{-0.25cm}
In this study, an RIS prototype has been utilized in a wireless communication system in order to increase the secrecy capacity between Bob and Eve. This objective is achieved by exploiting the propagation channel through configuring the phase shifts of the RIS elements. Both measurement and computer simulation results show that the secrecy capacity can be significantly improved by employing the RIS in an indoor environment. This improvement is obtained by assuming that the received power of Eve is available, which may not be true for practical scenarios. Although the initial results for providing PLS with RIS are promising, future work will investigate how these results can be achieved for real-life deployments. 

\section*{Acknowledgment}
We thank to StorAIge project that has received funding from the KDT Joint Undertaking (JU) under Grant Agreement No. 101007321. The JU receives support from the European Union’s Horizon 2020 research and innovation programme in France, Belgium, Czech Republic, Germany, Italy, Sweden, Switzerland, Türkiye, and National Authority TÜBİTAK with project ID 121N350. The work of E. Basar is also supported by TÜBİTAK under grant 120E401. 

% \balance
% % % % % % % % % % % % % % % % % % % % % % % % % % % %

\bibliographystyle{IEEEtran}
\bibliography{main.bib}
% % % % % % % % % % % % % % % % % % % % % % % % % % % %

\end{document}